\documentstyle[epsfig,nato]{crckapb} 

\def\beq{\begin{equation}}
\def\eeq{\end{equation}}

\def\bx{{\bf x}}
\def\br{{\bf r}}

\def\bv{{\bf v}}
\def\mpc{\,{\rm {Mpc}}}
\def\mpch{\,h^{-1}{\rm {Mpc}}}
\def\kms{\,{\rm {km\, s^{-1}}}}

\def\Deltae{\Delta_E}

\def\bxi{{\overline \xi}}
\def\vdisper{{\langle v^2_{12}(r)\rangle}}

\begin{opening}
\title {ANALYTIC APPROXIMATIONS TO GALAXY CLUSTERING}
\author {H.J. MO} 
\institute {Max-Planck-Institut f\"ur Astrophysik\\
Karl-Schwarzschild-Str. 1, 85748 Garching, Germany}
\end{opening}
\runningtitle{ANALYTIC APPROXIMATIONS TO GALAXY CLUSTERING}
\begin{document}
\begin{abstract} We discuss some recent progress in constructing 
analytic approximations to the galaxy clustering. We show that
successful models can be constructed for the clustering of both dark
matter and dark matter haloes. Our understanding of galaxy clustering 
and galaxy biasing can be greatly enhanced by these models. 
\end{abstract}

\section {Introduction}

The large scale structure of the Universe is believed to have developed
from small perturbations (usually assumed to be Gaussian) 
of the matter density field by gravitational instabilities. 
Under these assumptions the clustering pattern and velocity
field observed today are determined by the initial conditions via the 
perturbation power spectrum [$P(k)$] and the cosmological parameters such as  
$\Omega_0$, the cosmic density parameter. It is therefore possible  
to derive constraints on model parameters from the observed density and 
velocity distributions of galaxies. 

There are two fundamental problems 
to be addressed: First, since the galaxy distribution
may be biased relative to the mass density field, 
we need to understand such bias
before making meaningful comparisons between models and 
observations. Second, even if the observed 
galaxy distribution traces the matter distribution, we still need to  
understand how the observed distribution is related to a cosmogonical
model. This is by no 
means trivial, because the clustering pattern and velocity field 
observed today are nonlinear. N-body simulations are usually 
invoked to find a solution to these problems. However, such
simulations are limited both in resolution and in dynamical range,
and can be difficult to interpret. Our understanding of the underlying
physics can be greatly enhanced by simple physical models
and the analytic approximations they provide. 

In this article we summarize some of our recent progress in 
connection to the problems mentioned above.
We show that successful semianalytic
models can be constructed for the clustering
of both dark matter and dark matter haloes. Such models 
can enhance significantly our understanding of
both the nonlinear evolution of galaxy clustering 
and galaxy biasing. 

\section {Cosmogonies} 

The models present below are for CDM-like cosmogonies. 
The cosmology is described by the cosmological
matter density ($\Omega_0$), the cosmological constant ($\lambda_0$) 
and the Hubble constant ($H_0=100h\kms\mpc^{-1}$). 
The initial power spectrum is
\beq \label{one}
P(k)\propto {kT^2(k)},
\eeq
\beq
T(k)={{\rm ln}(1+2.34q)\over 2.34 q}
\left\lbrack 1+3.89 q+(16.1q)^2+(5.46q)^3+(6.71q)^4
\right\rbrack^{-1/4},
\eeq
where $q\equiv k/(\Gamma h {\mpc}^{-1})$, $\Gamma\equiv \Omega_0 h$
(Bardeen et al. 1986).
The {\rm RMS} mass fluctuation in top-hat windows with
radius $R$, $\sigma (R)$, is defined by
\beq
\sigma^2 (R)=\int_0^\infty {dk\over k}\Delta^2(k) W^2(kR),
\eeq
where $W(x)$ is the Fourier transform of the top-hat window function,
and 
\beq
\Delta^2(k)=
(1/ 2\pi^2) k^3P(k)
\eeq
is the power variance. We normalize $P(k)$ by specifying 
$\sigma_8\equiv \sigma(8\mpch)$.

\section {Low-Order Statistics of Dark Matter Distribution}   

\subsection {Mass Correlation Function}   

The {\it evolved}  two-point correlation function $\xi (r)$ is related 
to the {\it evolved} power variance $\Deltae^2 (k)$ by
\beq
\xi(r)=\int_0^\infty {dk\over k} \Deltae^2(k) {\sin kr\over kr}.
\eeq
Thus, in order to get $\xi(r)$ we need an expression for $\Deltae(k)$.
Following the original argument of Hamilton et al. (1991),
Jain, Mo \& White (1995),
Padmanabhan et al. (1996) and Peacock \& 
Dodds (1996, hereafter PD) have obtained fitting formulae which relate 
$\Deltae$ to $P(k)$ for a given cosmological model. 
The latest version of such a fitting formula is given in PD. 
The solid curve in Fig.1 shows the prediction of such
a formula for $\xi(r)$ in the standard cold dark matter
(SCDM) model with $\Omega_0=1$, $\lambda_0=0$, $h=0.5$ and
$\sigma_8=0.62$ (see Mo, Jing \& B\"orner 1997, hereafter MJB,
for details). Comparing it with the N-body results,
we see clearly that the fitting formula works well. 
This is true for various other cosmogonic models
studied.
\begin{figure}
\vspace{-1cm}  
\epsfig{figure=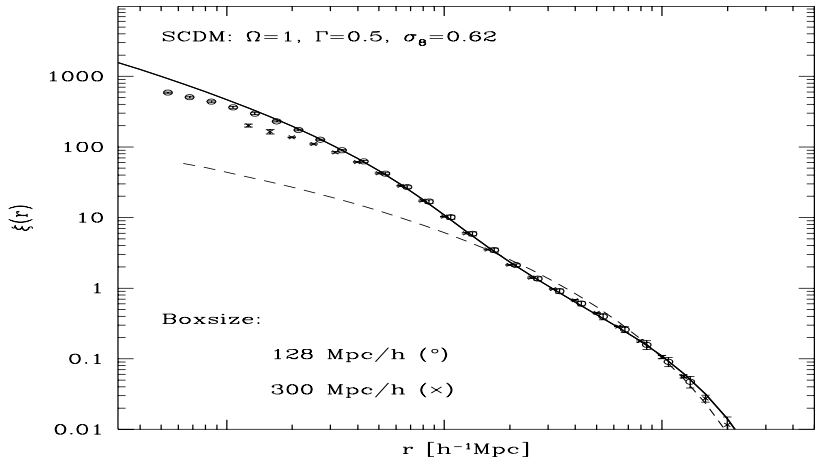,width=12.0cm,height=8.0cm}
\vspace{-1cm}  
\caption{Predicted two-point correlation function of mass
(solid curve) compared with simulation results (symbols).
The dashed curve is given by the linear power spectrum.}
\end{figure}
\begin{figure}
\vspace{-1cm}  
\epsfig{figure=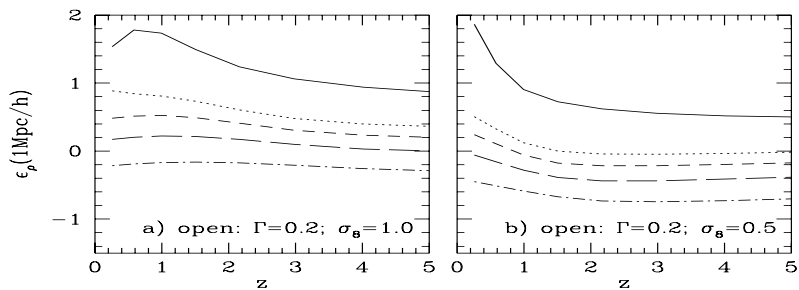,width=12.0cm,height=9.0cm}
\vspace{-4cm}  
\caption{The redshift-evolution parameter of the mass correlation
function as a function of redshift, for open cosmologies
with $\Omega_0=0.1$, 0.2, 0.3, 0.5 and 1 (from bottom up).}
\end{figure}
   
\subsection{Redshift Evolution of Mass Correlation Function} 

 The redshift evolution of the two-point
correlation function is usually parametrized by the form
\beq
\xi(r,z)=\xi(r,0)(1+z)^{-(3+\epsilon_{\rho})},
\eeq
where $\xi(r,z)$ is the amplitude of the two-point correlation
function at {\it physical} radius $r$ at redshift $z$, 
$\epsilon_{\rho}$ is a parameter to describe the time evolution.
If $\xi(r)\propto r^{-\gamma}$, then $\epsilon_{\rho}=\gamma-1$ for 
the linear growth in an Einstein-de Sitter universe,
$\epsilon_{\rho}=3-\gamma$ for clustering patterns fixed in comoving space,
and $\epsilon_{\rho}=0$ for stable clustering (i.e. clustering patterns
fixed in proper coordinates). Given the model in \S 3.1, it is
straightforward to obtain $\epsilon_{\rho}$. Fig. 2 
shows $\epsilon_{\rho}$ as a
function of $z$ for open cosmologies with various $\Omega_0$. 
The value of $\epsilon_{\rho}$ at $z$ is obtained by fitting
$\xi(r=1h^{-1}{\rm Mpc}, z')$ in the redshift interval,
$z'=0\to z$. The evolution is more rapid (i.e. $\epsilon_{\rho}$ larger)
for universes with larger $\Omega_0$, since linear structures 
grow faster in a high-$\Omega_0$ universe. For high $\Omega_0$,
the evolution is faster at $z\sim 1$, because of nonlinear evolution.
At late time when clustering on $r\sim 1h^{-1}{\rm Mpc}$ becomes
stable, the evolution becomes slower.      
\begin{figure}
\vspace{-1cm}  
\epsfig{figure=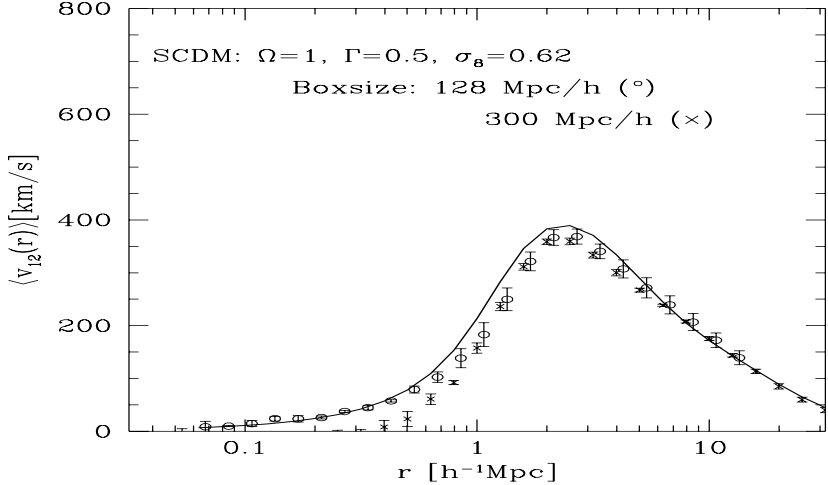,width=12.0cm,height=8.0cm}
\vspace{-1cm}  
\caption{Mean pairwise peculiar velocity of dark matter
particles. Curve is the model prediction; symbols
are the simulation results.}
\end{figure}

\subsection {Mean Pairwise Peculiar Velocities}

   From the pair conservation equation (Peebles 1980, \S 71),
the ensemble (pair weighted) average of the pairwise peculiar
velocity $\langle v_{12}(r)\rangle 
\equiv \langle [\bv(\bx)-\bv(\bx+\br)]\cdot
{\hat {\br}}\rangle$ can be written as
\beq
{\langle v_{12}(r)\rangle \over H(a)r}=-{1\over 3} {1\over [1+\xi(y,a)]}
{\partial \bxi (y,a) \over \partial \ln a},
\eeq
where $r$ is the proper, and $y$ the comoving, separation  
between the pairs;
$H$ is the Hubble's constant at expansion factor $a$;
$\bxi(y,a)\equiv (3/ y^3)\int_0^y y ^2dy  \xi(y ,a)$.
Thus, to obtain $\langle v_{12}(r)\rangle$, we need to work out 
$\partial \Deltae(k,a)/\partial a$. This can be done
directly from the fitting formula of $\Deltae(k,a)$
(see MJB). Fig. 3 shows the comparison between the model prediction
and the simulation results for the SCDM model.  
The agreement between the two is remarkably good, and
this is true for many other cosmogonic models studied.

\subsection {Cosmic Energy Equation} 

The (density weighted) mean square peculiar velocity
of mass particles $\langle v_1^2\rangle $ is related to the two-point 
correlation function by the cosmic energy equation: 
\beq\label{x.8}
{d\over da} a^2\langle v_1^2\rangle =4\pi G{\overline \rho} a^3
{\partial I_2(a)\over \partial \ln a}, 
\eeq
where ${\overline \rho}$ is the mean density of the universe, and 
\beq
I_2(a)\equiv \int _0^\infty ydy\xi(y,a)=\int_0^\infty {dk\over k}
{\Deltae^2(k,a)\over k^2}.
\eeq
Integrating eqn. (\ref{x.8}) once, we have
\beq\label{x.10}
\langle v_1^2\rangle ={3\over 2}\Omega (a) H^2(a)a^2I_2(a)
\left\lbrack 1-{1\over a I_2(a)}\int_0^a I_2(a )da \right\rbrack.
\eeq
In the linear case, $I_2(a)\propto D^2(a)$, where 
$D(a)=ag(a)$ [$g(a)$ is the linear growth factor; $a_0=1$] 
and
\beq\label{x.11}
\langle v_1^2\rangle ={3\over 2}\Omega (a) H^2(a) a^2 I_2(a)
\left\lbrack 1-{1\over aD^2(a)}\int_0^a  D^2(a )da
\right\rbrack.
\eeq
We found that eqn.(\ref{x.11}) is a good approximation 
(to an error of $<10\%$) to eqn.(\ref{x.10}) for all realistic 
cases. Thus, for a given cosmogonic model, we can easily obtain
$\langle v_1^2\rangle $. For various cosmogonies, the model
predictions fit the simulation results to an accuracy better
than 10 percent (see MJB). 

\subsection {Pairwise Peculiar Velocity Dispersion}

The relative velocity dispersion of particle pairs
of separation $r$ is defined as 
$\langle [\bv (\bx)-\bv (\bx+\br)]^2\rangle ^{1/2}$.
In Fig. 4 we show (by symbols) the dispersion of the 
pairwise peculiar velocities projected along the separations
of particle pairs
[$\langle v_{12}^2(r)\rangle^{1/2}$] in the N-body simulations. 
The main features of $\langle v_{12}^2(r)\rangle^{1/2}$
are (a) monotonic rise at small $r$; (b) saturation
at large $r$; (c) a maximum at medium $r$. 
As shown in MJB, these features can all be explained 
by physical arguments.  
\begin{figure}
\vspace{-1cm}  
\epsfig{figure=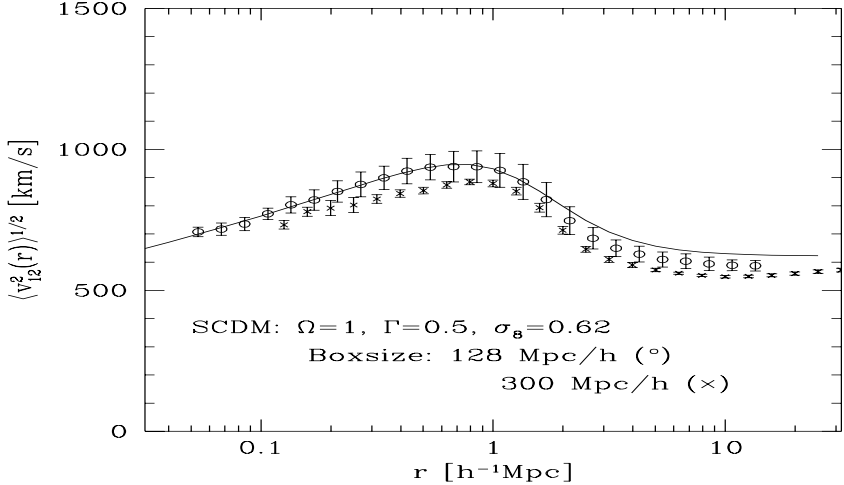,width=12.0cm,height=8.0cm}
\vspace{-1cm}  
\caption{Pairwise peculiar velocity dispersion of dark matter
particles. Curve is the model prediction; symbols are the
simulation results.}
\end{figure}

Based on the N-body results we make the following 
ansatz for $\vdisper$:
\beq
\vdisper ^{1/2}=\Omega^{0.5}Hr_c\phi(r/r_c),
\eeq
where $\phi(x)$ is a universal function and $r_c$ is a nonlinear
scale. We choose $r_c$ to be the virial
radius of $M^*$ haloes, $r_v^*$. For a given power spectrum,
the linear radius of $M^*$ haloes, $r_0^*$ , is given 
by $\sigma (r_0^*)=1$. The relation between $r_v^*$ and 
$r_0^*$ in different cosmological models can be 
obtained analytically, as discussed in detail in MJB.
We approximate the functional form
of $\phi (x)$ by:
\beq
\phi(x)={\phi_\infty (1+Bx^{-\beta})+Ax^{-\alpha}
\over 1+V[g(a)]^{0.35}
[\Omega(a)]^{0.2}x^{-(\alpha+\kappa)}} ,
\eeq
where $\phi_\infty 
=\sqrt{{2\over3}}\langle v_1^2\rangle^{1/2}
/(Hr_c\Omega^{0.5})$; $A$, $B$, $V$, 
$\alpha>\beta>0$, and $\kappa>0$ are constant.
$\vdisper$ is forced to have the 
large separation asymptotic value for uncorrelated pairs. 
For $r\to 0$, $\vdisper ^{1/2}\propto x^{\kappa}$
so that it increases with $r$ as a power law. 
The solid curve in Fig. 4 shows the prediction of our
fitting formula with
\beq
A=58.67;\,\,\, B=-0.3770;\,\,\, V=4.434;
\eeq
\beq
\alpha=2.25;\,\,\, \beta=1.90;\,\,\, \kappa=0.15 .
\eeq
The fit to the simulation data is reasonably good.
   
\section {Spatial Clustering of Dark Matter Haloes}

 So far we have discussed the clustering properties of dark matter.
To compare model predictions with the observed galaxy distribution,
we also need to understand how the galaxy distribution is related
to the dark matter distribution. The bias of the galaxy distribution 
relative to the mass distribution can be obtained once we know how
galaxies form in the mass density field. However, some progress 
can still be made before the details of galaxy formation is 
understood. In the standard scenario of galaxy formation, a 
gravitationally dominant dissipationless component of dark matter
is assumed to aggregate into dark matter clumps (dark matter haloes),
galaxies then form by the cooling and condensation of gas within these
dark haloes (e.g. Kauffmann 1997). It is therefore important 
to approach the problem of galaxy biasing by first understanding
how dark matter haloes are distributed relative to the mass.
In the following we show that simple analytic models for such
relations can be constructed (see Mo \& White 1996, MW).

\subsection {Bias Relation}

  We define dark matter haloes as virialized clumps of dark matter.
In the spherical collapse model, a dark matter halo is 
characterized by its mass, $M$,  and the redshift, $z$, 
when it is assembled. To describe the relation between
halo and mass distributions we define a bias relation, 
\beq\label{y.1}
\delta_h(R)=b(R,\delta, M, z)\delta(R),
\eeq
where $\delta(R)=[\rho(R)-{\overline \rho}]/ 
{\overline \rho}$ is the overdensity of matter in a sphere
of radius $R$, $\delta_h (R)$ is the same overdensity for
dark matter haloes. Suppose the conditional density
of dark matter haloes within a sphere of radius $R$
is $n_h(M,z\vert \delta,R)$. 
(This is the number density of dark matter
haloes within a sphere of radius $R$, given that the 
mean mass overdensity within this sphere is $\delta$.) The bias
relation can then be written as  
\beq
\delta_h(M,z\vert \delta, R)={{n}_h(M,z\vert \delta,R)\over 
{\overline n}_h(M,z)}-1,
\eeq
where ${\overline n}_h(M,z)$ is the mean number density of haloes
in the universe, given by the Press-Schechter formalism 
(Press \& Schechter 1974, PS):
\beq
{\overline n}_h(M,z)dM=-\sqrt{{2\over \pi}}{{\overline \rho}\over M}
{\delta_z\over \sigma^2(r)} {d\sigma\over dM}
\exp\left[-{\delta_z^2\over 2\sigma^2(r)}\right] dM,
\eeq
where $M=(4\pi/3){\overline \rho} r^3$, $\delta_z=\delta_cD(a_0)/D(a)$,
$\delta_c\approx 1.686$. 
The conditional density, $n_h(M,z\vert \delta,R)$,
can be obtained by an extension of the PS formalism 
(e.g. Bower 1991):
\begin{eqnarray}
n_h(M,z\vert \delta_0, R_0)dM &=&
-\sqrt{{2\over \pi}}{{\overline \rho}\over M}
{(\delta_z-\delta_0)\sigma(r)\over[\sigma^2(r)-\sigma^2(R_0)]^{3/2}} 
{d\sigma(r)\over dM} \nonumber \\
&\times &\exp\left\{-{(\delta_z-\delta_0)^2\over
2[\sigma^2(r)-\sigma^2(R_0)]}\right\} 
dM,
\end{eqnarray}
where $\delta_0$ is the linear mass 
overdensity of spherical regions with Lagrangian radius $R_0$. 
Assuming spherical collapse, we have $R_0=[1+\delta (R)]^{1/3}R$, 
and $\delta_0$ is determined by $\delta (R)$ and $R$
through the spherical collapse model (see MW for details). 
With these the bias relation is fixed. 
Fig. 5 shows the bias
relation for haloes in a scale free model with $P(k)\propto k^{-0.5}$.
It is clear that our simple model works reasonably well.
\begin{figure}
\vspace{-0.5cm}  
\epsfig{figure=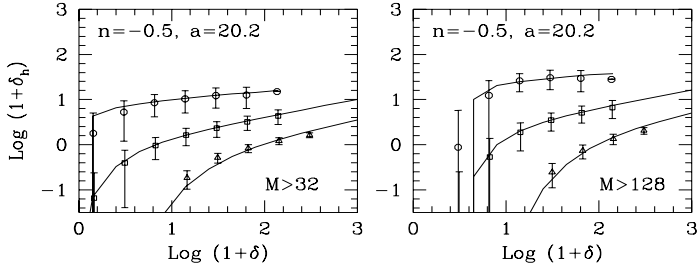,width=12.0cm,height=12.0cm}
\vspace{-8cm}  
\caption{Bias relation $\delta_h(\delta)$ in a scale 
free model with $n=-0.5$. Results are shown for spheres with
radii $R/L=0.02$, 0.05 and 0.13 ($L$: the side of the simulation
box). The results for $R/L=0.05$ and 0.13 are shifted by 1 and 2
decades along the horizontal axis.}
\end{figure}
\begin{figure}
\vspace{-0.5cm}  
\epsfig{figure=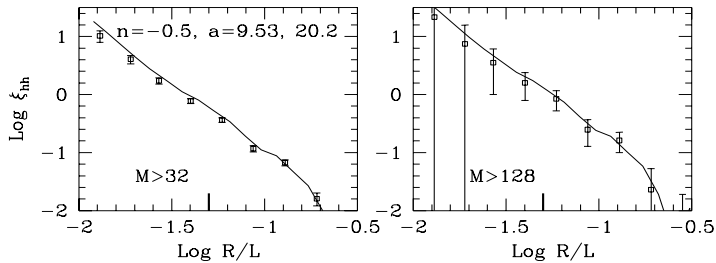,width=12.0cm,height=12.0cm}
\vspace{-8cm}  
\caption{Predicted halo-halo two-point correlation functions
(solid curves) compared to those from simulations (symbols).}
\end{figure}

\subsection {Two-Point Correlation Function of Haloes}

 When $\delta\ll 1$ and $r\ll R_0$, the bias relation can be written as
\beq
\delta_h(R; M,z)=b(M,z)\delta(R);\,\,\,\,
b(M,z)=1+{1\over \delta_z}
\left[{\delta_z^2\over \sigma^2(r)}-1\right].
\eeq
Thus the bias factor $b$ depends only on $M$ and $z$.
Under the assumption of linear bias, 
\beq
\xi_h(r)=b^2(M,z)\xi(r),
\eeq
where $\xi(r)$ is the two-point correlation function of mass. 
Fig. 6 shows the two-point correlation functions
for haloes in a scale free model with $P(k)\propto k^{-0.5}$.
The simple model works reasonably well.

\begin{figure}
\vspace{-1cm}  
\epsfig{figure=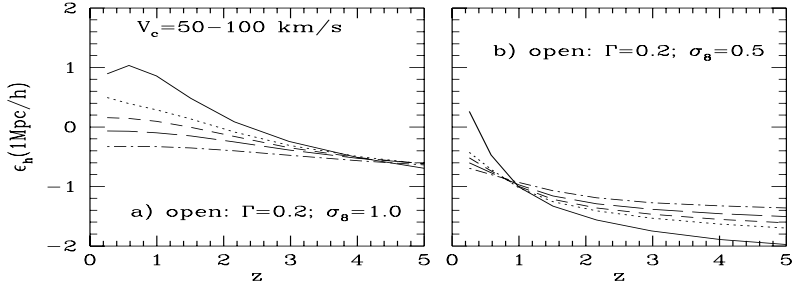,width=12.0cm,height=9.0cm}
\vspace{-4.5cm}  
\caption{The redshift-evolution parameter of the halo-halo
correlation function for haloes selected according to the 
first selection rule (see text). Results are shown 
for open cosmologies with $\Omega_0=0.1$, 0.2, 0.3, 0.5 and 1
(from dot-dashed to solid curves).}
\end{figure}
\begin{figure}
\vspace{-1cm}  
\epsfig{figure=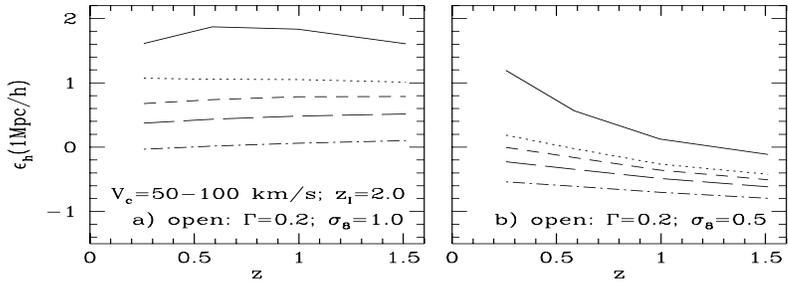,width=12.0cm,height=9.0cm}
\vspace{-4.5cm}  
\caption{The same as Fig. 7 for haloes selected according to the
second selection rule.}
\end{figure}

\subsection {Redshift Evolution of Halo Correlation Function}

 As in \S 3.2, we parametrize the redshift evolution of
$\xi_h$ by 
\beq  
\xi_h(r,z)=\xi_h(r,0)(1+z)^{-(3+\epsilon_h)}.
\eeq
Unlike the mass correlation function $\xi$, the evolution
of $\xi_h$ depends on how dark matter haloes are 
selected. We consider two different selections.
In the first, haloes are selected at the same time when the 
correlation function is calculated. This case is relevant for
galaxies, if galaxies merge as fast as their dark haloes.
In the second, haloes are selected at an earliear epoch
than when their correlation function is calculated.
This case is relevant for galaxies, if they remain distinct
after their haloes merge. Fig. 7 shows $\epsilon_h$ as a
function of $z$ in open cosmologies with two
choices of $\sigma_8$. Here haloes are selected according to the
first selection. As before,
the value of $\epsilon_h$ at $z$ is obtained by fitting
$\xi_h(r=1h^{-1}{\rm Mpc}, z')$ in the redshift interval $z'=0\to z$. 
In all cases, the halo correlation function evolves less rapidly
than the mass correlation function (see Fig. 2), 
because haloes with fixed circular velocities are more
biased at higher redshifts. The evolution is also
less rapid for a lower $\sigma_8$, because of the higher
degree of bias involved. Fig. 8 shows the same results 
for haloes selected according to the second selection.
In this case, the evolution of $\xi_h$ is faster 
than that of $\xi$, because the haloes (with $V_c=50-100\kms$)
are antibiased relative to the mass at the time of selection
($z=2$). For massive haloes which are ({\it positively}) 
biased relative to the 
mass, the evolution of their correlation function is slower
than that of the mass. The evolution is also slower for a lower
$\sigma_8$, because of the increased degree of bias (or
decreased degree of antibias). 
The results in this subsection suggest that one needs 
to be very cautious when interpreting the redshift evolution of galaxy
correlation function. Without knowing in detail 
what population the observed galaxies are, it is difficult to 
infer the time evolution of the mass correlation from
the correlation functions of these galaxies.

\section {Discussion}

  The models presented above show how the low order statistics 
of the density and velocity distributions are determined by
cosmogonies. Thus they can be used to construct statistical
measures of the density and peculiar velocity fields
to constrain cosmogonic models by observations. In particular,
the models can help in the reconstruction of cosmogonic 
parameters from measurements in redshift space. There are
also other applications. MJB discuss the dependence 
of the small scale pairwise peculiar velocity dispersion on the 
presence (or absence) of rich clusters of galaxies.
Mo, Jing \& White (1996) use the model to study the
correlation of galaxy clusters. Mo, Jing \& White (1997) extend
the model to high-order correlations of dark matter haloes
and density peaks.

\end{document}